\documentclass{optica-article}

\journal{opticajournal}

\articletype{Research Article}

\usepackage{lineno}
\usepackage{physics}
\usepackage{arydshln}
\usepackage{float}

\begin{document}

\title{High-speed integrated QKD system}

\author{Rebecka Sax\authormark{1,*}, Alberto Boaron\authormark{1}, Gianluca Boso\authormark{1,2}, Simone Atzeni\authormark{3,4}, Andrea Crespi\authormark{3,4} Fadri Grünenfelder\authormark{1}, Davide Rusca\authormark{1}, Aws Al-Saadi\authormark{5}, Danilo Bronzi\authormark{5}, Sebastian Kupijai\authormark{5}, Hanjo Rhee\authormark{5}, Roberto Osellame\authormark{3,4} and Hugo Zbinden\authormark{1}}

\address{\authormark{1}Group of Applied Physics, University of Geneva, Rue de l'Ecole-de-Médecine 20, 1205, Genève, Switzerland\\
\authormark{2}ID Quantique SA, Rue Eugène-Marziano 25, 1227, Genève, Switzerland\\
\authormark{3}Institute for Photonics and Nanotechnology, CNR-IFN, 20133, Piazza Leonardo da Vinci, Milano, Italy\\
\authormark{4}Dipartimento di Fisica, Politecnico di Milano, 20133, Milano, Italy\\
\authormark{5}Sicoya GmbH, Carl-Scheele-Strasse 16, 12489, Berlin, Germany
}
\email{\authormark{*}rebecka.sax@unige.ch}

\begin{abstract}
Quantum key distribution (QKD) is nowadays a well established method for generating secret keys at a distance in an information-theoretic secure way, 
as the secrecy of QKD relies on the laws of quantum physics and not computational complexity. In order to industrialize QKD, low-cost, mass-manufactured and practical QKD setups are required. Hence, photonic and electronic integration of the sender's and receiver's respective components is currently in the spotlight. Here we present a high-speed (2.5~GHz) integrated QKD setup featuring a transmitter chip in silicon photonics allowing for high-speed modulation and accurate state preparation, as well as a polarization-independent low-loss receiver chip in aluminum borosilicate glass fabricated by the femtosecond laser micromachining technique. Our system achieves raw bit error rates, quantum bit error rates and secret key rates equivalent to a much more complex state-of-the-art setup based on discrete components \cite{Boaron2018, Boaron_2018_fast_simple}. 
\end{abstract}



\section{Introduction}

The security of the exchange of an encrypted message is an extremely relevant issue in today's society, as disastrous consequences can arise when it is compromised. One up-and-rising threat is the quantum computer, which would be able to efficiently crack the current most-used encrypting techniques \cite{Shor_1997} and whose technology matures as the author is writing this article \cite{Mirko_IBMQ, Wright_QC}. Hence the natural entry of quantum key distribution (QKD), which establishes an information-theoretically secure key exchange, providing long term security.  

Since the first proposal of a QKD protocol in 1984 \cite{Bennett1984} and its first experimental realization in 1992 \cite{Bennett92experimentalquantum}, more protocols and a multitude of experiments have been established. This global enthusiasm has resulted in enormous increases in the communication distance (using fiber \cite{Boaron2018, Lucamarini2018, Wang2022}, as well as free-space \cite{Liao2017}) and in the secret key rate \cite{yuan2018, Fadri_High_Speed}. 

In order to industrialize QKD and to merge it with existing networks, a vision of integrated transmitters and receivers separated at metropolitan distances seems rather judicious. The miniaturization of such systems is notably important, with advantages in terms of low cost, mass production, scalability, simple stabilization in temperature and compatibility with CMOS-production.

The first realization of a fully integrated QKD system (both the transmitter and receiver integrated) consisted of a silicon transmitter and a SiO$_x$N$_y$ receiver operating at 560~MHz clock rate, using the BB84 time-bin protocol at 20~km distance separation \cite{Sibson2017_2}. Subsequently, several integrated implementations have been reported for various QKD schemes \cite{Sibson2017, Paraiso2021, Paraiso2019, Ma:16, Bunandar18, Kong:20, Geng2019, Semenenko:20, Wei2020, Vest_2022}. Some included an integrated laser \cite{Sibson2017, Paraiso2021, Paraiso2019} and other presented hybrid versions that maintain one of the components as non-integrated (either the transmitter or the receiver device) \cite{Paraiso2021, Paraiso2019, Bunandar18, Ma:16}. Integrated detectors on chip have also been realized \cite{beutel2021detector}. 


Here we present a 2.5~GHz integrated QKD system, the fastest integrated system to our knowledge \cite{Liu_review}, which features a precise state preparation and a polarization independent receiver. At a distance of 151.5~km of standard single-mode fiber we obtain a secret key rate (SKR) of 1.3~kbps using InGaAs/InP negative feedback avalanche photodiodes. We further demonstrate extremely low quantum bit error rates (QBERs) (QBER$_z$ of 0.9\% and $\phi_z$ of 2.2\%) using superconducting nanowire single-photon detectors (SNSPDs) at a distance of 202.0~km, thereupon raising the bar of the state-of-the-art integrated QKD and further laying the groundwork for its use.


\section{QKD protocol}
\begin{figure}[htbp]
\centering
\includegraphics[width=0.6\linewidth,trim={0 0 2.1cm 0}, clip]{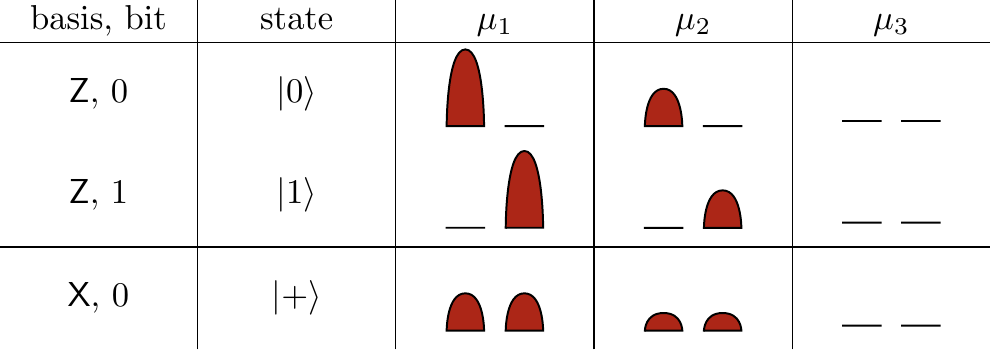}
\caption{Encoding of the states sent by Alice \cite{Boaron_2018_fast_simple}. $\mu_1$ and $\mu_2$ correspond to the two mean photon numbers used for the one-decoy state protocol.}\label{Encoding_Alice}
\end{figure}

We apply a 3-state BB84 protocol using the one-decoy state method \cite{Boaron_2018_fast_simple, Rusca_2018}, with time-bin encoding. The three states, and their respective decoys, prepared by Alice are shown in figure \ref{Encoding_Alice}. They belong to one of the two bases, Z and X, and they are chosen at random. The two states in the Z basis are: 

\begin{equation}
\ket{0} = \ket{\alpha}_E\ket{0}_L,
\end{equation}
\begin{equation}
\ket{1} = \ket{0}_E\ket{\alpha}_L.
\end{equation}
$E$ standing for \textit{early}, $L$ for \textit{late} and $\ket{\alpha}$ for a weak coherent state. The state in the X basis is 
\begin{equation}
\ket{+} = \frac{1}{\sqrt{2}}(\ket{0} + \ket{1}).
\end{equation}

Qubits detected in the Z basis will undergo a time-of-arrival measurement and constitute the raw key. Qubits detected in the X basis will pass through an imbalanced Mach-Zehnder interferometer (imb-MZI) and corresponding detections will reflect the security of the exchange of the secret key. 

\section{Experimental setup}\label{exp_setup}
\begin{figure}[H]
\centering
\includegraphics[width=0.8\linewidth]{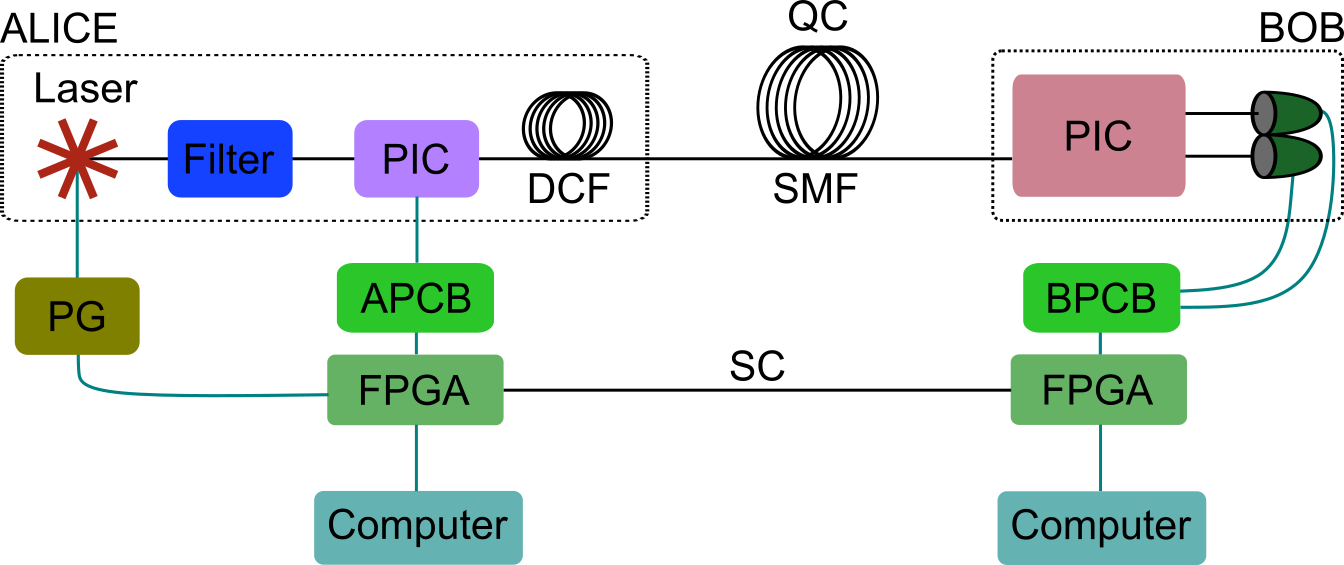}
\caption{Simplified schematics of the experimental setup. PIC = photonic integrated circuit, DCF = dispersion compensating fiber, QC = quantum channel, SMF = single-mode fiber, PG = pulse generator, APCB = Alice printed circuit board, BPCB = Bob PCB, FPGA = field-programmable gate array, SC = service channel. Black lines correspond to optical links and blue lines correspond to electrical connections.}\label{Full_setup}
\end{figure}

An overview of the full QKD setup is depicted in figure \ref{Full_setup}. Alice, the transmitter, and Bob, the receiver, are connected via a quantum channel (QC) and a service channel (SC). The former serves for guiding the quantum encoded states and the latter for classic (public, but authenticated) communication between the parties. Each of the two apparatuses is controlled by a field-programmable gate array (FPGA), which also allows for synchronization and communication of the two parties, via the SC.  

Regarding the optical elements, the transmitter encompasses a distributed feedback (DFB) laser with a filter, a photonic integrated circuit (PIC) and a dispersion compensating fiber (DCF). Phase-randomized pulses of light at a repetition rate of 2.5~GHz and FWHM of around 31~ps are generated by a gain-switched high-bandwidth DFB laser at 1550~nm (Gooch and Housego). The pulse train enters the integrated transmitter chip where the three states and their decoys are produced at random using the following components: imb-MZI, intensity modulator (IM) and variable optical attenuators (VOA). The probability to select the basis Z ($p_z$) and X ($p_x$) is 0.67 and 0.33, respectively. The random numbers used to choose the states are produced by AES (Advanced Encryption Standard) cores seeded by a Quantis Quantum Random Number Generator (QRNG) from ID Quantique SA. Upon exiting the chip, light pulses travel through the DCF to pre-compensate for all the chromatic dispersion created on the trip from Alice to Bob in the QC. The QC consists of standard single-mode fibers (SMF) with around 0.2~dB/km losses.

On the receiver side, the integrated part consists of a passive beam splitter and an imb-MZI. The effective splitting ratio for the Z and X basis, i.e. taking into account different losses in respective optical paths, is 94/6. The imbalance of the interferometer of Bob should be ideally the same as that of Alice, i.e. 200~ps. However, due to fabrication uncertainties, a delay difference of around 1.6~ps between the two interferometers is measured. The main effect of a delay difference is on the QBER in the X basis, QBER$_x$, as it leads to a reduced interference of the pulses in the imb-MZI. The relative phase of their interferometers is actively adjusted by acting on the phase of Alice's interferometer in such a way that the two pulses interfere destructively in the output we monitor in the X basis. A feedback loop is locked to minimize the number of detections in this output. It should be noted that since the occurrences are already low, the active adjustment will be more difficult with increased channel loss due to the, at that point, even lower statistics. The second output of the imb-MZI on the receiver side is not monitored.  

The (off-chip) single-photon detectors adopted for our main experiment are InGaAs/InP negative feedback single photon avalanche diode (SPADs) cooled by a free-piston Stirling cooler to -85$^{\circ}$C \cite{Korzh_2014}. For the characterization of our system  we also use in-house superconducting-nanowire single-photon detectors (SNSPDs) cooled at 0.8~K \cite{Caloz_2018}. These detectors feature low timing jitter (around 40~ps), negligible after-pulsing probability, high detector efficiency (around 80~\%) and low dark count rates ($dc_z = 200$, $dc_x = 100$). The SPADs are used for the experiment as these detectors are more mature than SNSPDs for practical real-world applications. It should be noted that the fixed 94/6 splitting ratio of the integrated beamsplitter on Bob's chip is suited for intermediate distances in this proof-of-principle experiment. Indeed, for short distances the large number of photons in the Z basis would rapidly saturate the SPADs, whereas for long distances too few detections in the X basis would give rise to non-negligible dark count contribution. However, versatility of the system could be easily increased by replacing the passive beam splitter at the receiver side with a tunable Mach-Zehnder interferometer.

\section{Integrated Transmitter}
Several challenges arise in the realization of integrated systems for QKD purposes depending on the protocol one uses. For what we are concerned, due to our high clock rate, we need accurate modulation of the quantum states at high frequencies on the transmitter side. Indeed, accuracy reflects on the extinction ratio (ER) of the quantum states and finally on the QBER. Moreover, for time-bin encoding, the platform must allow for the implementation of a MZI with high imbalance. 

\begin{figure}[htbp]
\centering
\includegraphics[width=0.5\linewidth]{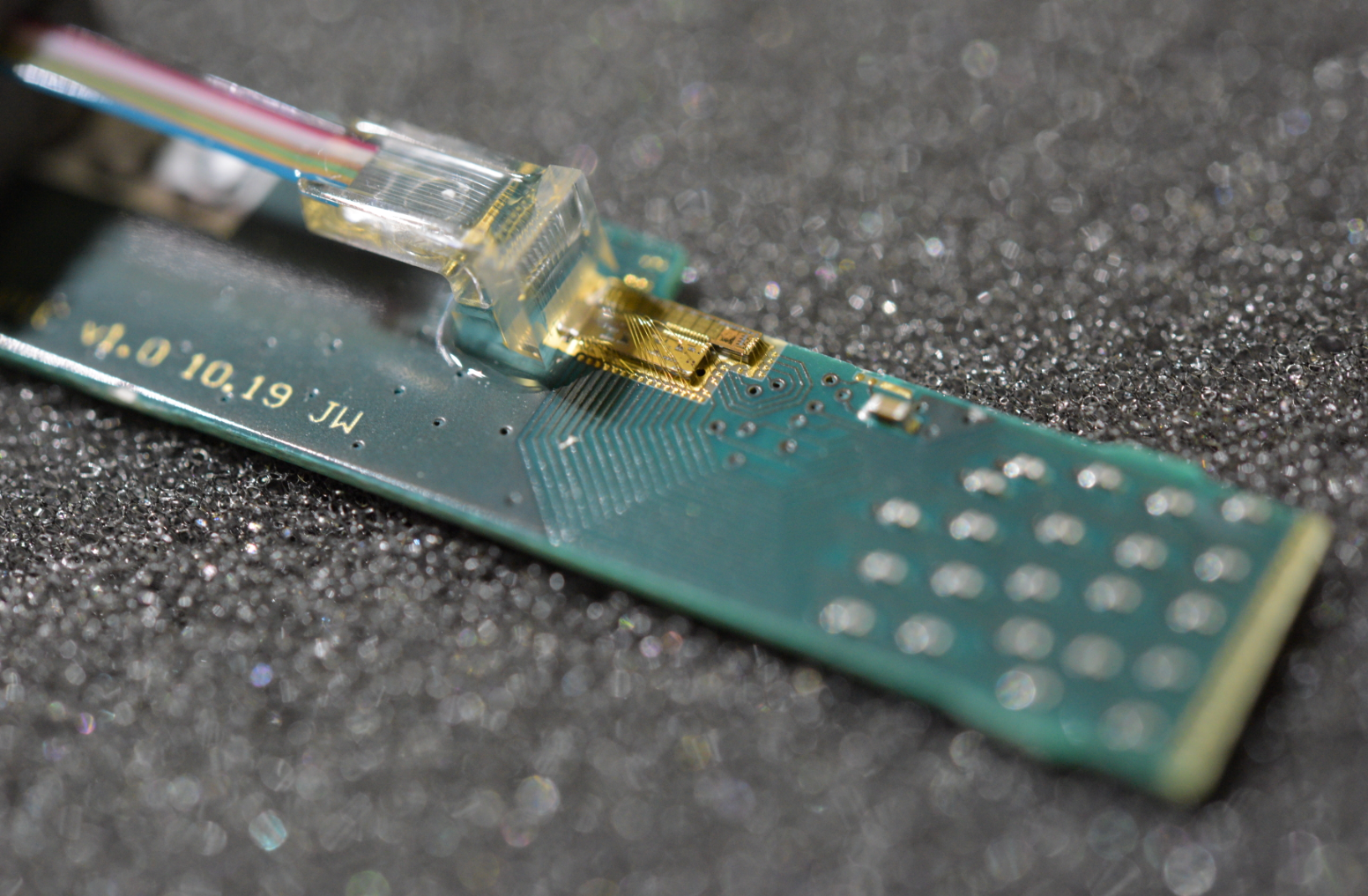}
\caption{Photo of the transmitter integrated circuit.}\label{Photo_Alice}
\end{figure}

We developed an integrated chipset based on silicon photonics, with the formerly mentioned qualities for the transmitter, in collaboration with Sicoya GmbH. It consists of a photonic integrated circuit (PIC), which is as small as 4.5~mm $\times$ 1.1~mm and an adjacent electronic driver integrated circuit (EIC)  0.75~mm $\times$ 1.1~mm, see figure \ref{Photo_Alice}. It is highly advantageous to use silicon photonics for our system as now most of the expensive electronics are on-chip. Additionally, it allows for high component density and small footprints. As can be seen in figure \ref{Photo_Alice}, the ICs are glued on and bonded to a small printed circuit board (PCB). This PCB is combined with a larger one and connected to a computer-controlled FPGA, as shown in figure \ref{Full_setup}. Light is coupled to the PIC via a fibre array and grating couplers. The chip is temperature stabilized at 45°C using a standard Peltier cooler/heater placed under the host PCB of the PIC. 

The chips were fabricated in the 0.13 µm SG25PIC SiGe bipolar-complementary metal-oxide-semiconductor (BiCMOS) process at the Leibniz Institute for High Performance Microelectronics (IHP) in Frankfurt (Oder), Germany, using 200~mm silicon-on-insulator (SOI) wafers and 248~nm Deep Ultra-Violet (DUV) lithography \cite{Website_IHP}. The nanowires are embedded within the 220~nm thick silicon device layer of the SOI substrate. The SOI rib-waveguides have dimensions of 220 $\times$ 450~nm$^2$ and are fabricated in a shallow trench process. The etching depth of the photonic structures is 170~nm, with a 50~nm high remaining slab on top of the underlying SiO$_2$ BOX-layer with a thickness of 2~$\mu$m. The implant doping level inside the p$^+$- and n$^+$-doped regions of the electro-optic phase shifters (EOPS) is 1$\cdot10^{20}$~cm$^{-3}$. The process provides a CMOS back-end-of-line with a stack of five metal layers. For fabrication of the driver chips, the SG25H4 SiGe BiCMOS technology also from IHP was used.

\begin{figure}[H]
\centering
\includegraphics[width=0.8\linewidth]{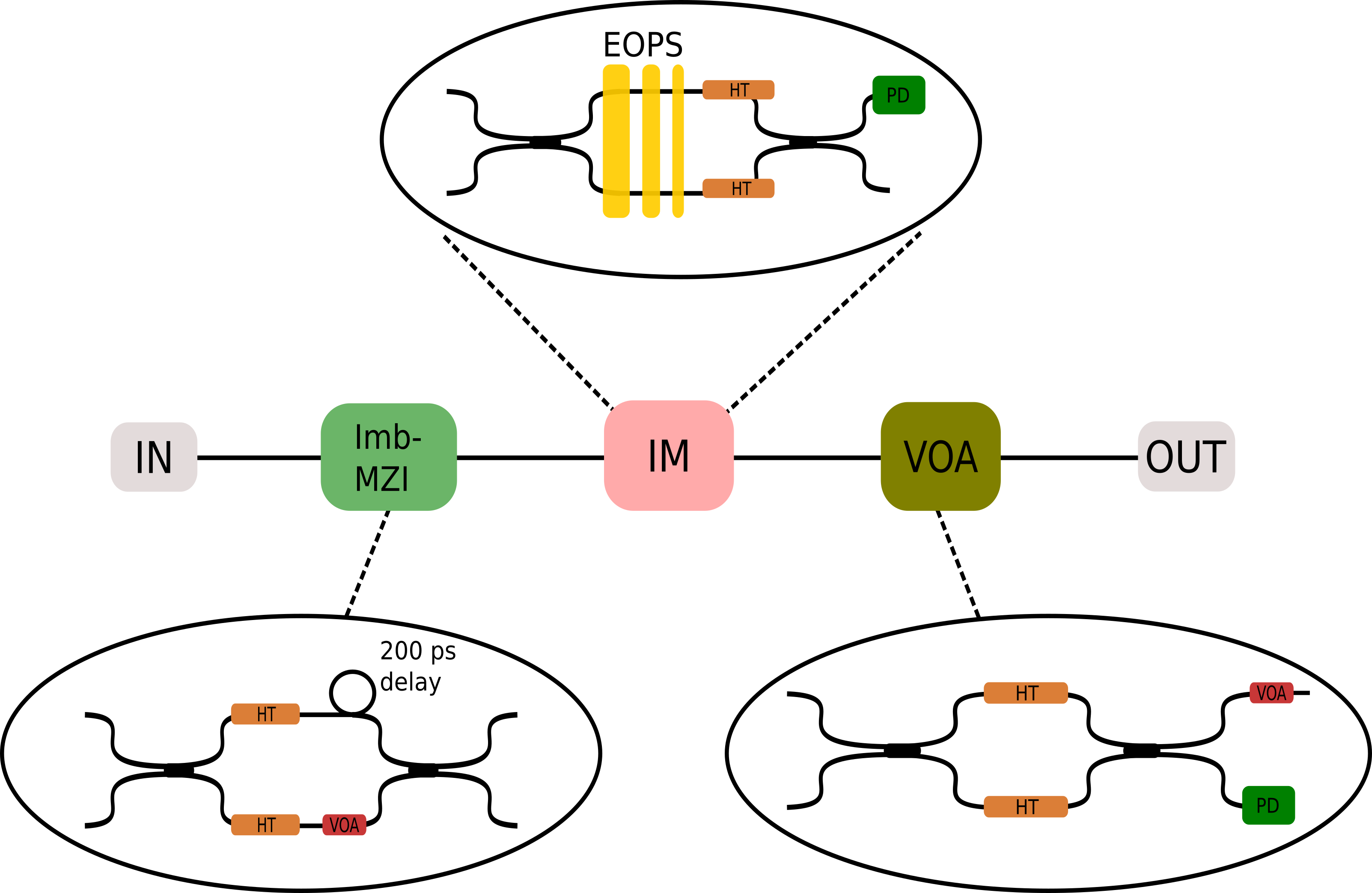}
\caption{Structure of the integrated transmitter circuit. Imb-MZI = imbalanced Mach-Zehnder Interferometer, IM = intensity modulator, VOA = variable optical attenuator, HT = heater, EOPS = electro-optic phase shifter, PD = photodiode.}\label{Structure_Alice}
\end{figure}

Figure \ref{Structure_Alice} reports a functional scheme of the transmitter device. Light entering the PIC passes first through an imb-MZI. The phase of the interferometer can be controlled via thermo-optic phase shifters (TOPSs or heaters), one in each arm, one of which is adjusted for the active phase stabilization between Alice and Bob. The shorter arm also comprises a VOA (based on carrier absorption) to compensate for propagation loss in the longer arm. Light then enters an IM based on a balanced Mach-Zehnder-Modulator (MZM). In the arms of the IM there are three EOPSs based on carrier injection. Each EOPS has been fabricated with a specific size and is designed for a given amplitude of modulation. In addition, each EOPS is connected to the analog driver circuit on the EIC via wire-bondings. This allows to individually actuate each EOPS and produce the full combination of quantum states. Likewise to the imb-MZI, the two arms of the IM include heaters, used to adjust its working point. The driver chip is a digitally Serial Peripheral Interface (SPI) controlled driver with limiting amplifier for high bandwidth and high voltage swing applications with a 3.3~V power supply. It consists of three active stages: a limiting amplifier, a buffer stage, and a current-mode logic (CML) output driver, as well as an active input matching network consisting of two common-base transistors. The core of the single-channel driver is very small, and the entire layout of the cell circuitry was kept strictly, thermally and electrically, symmetric with respect to the radio-frequency (RF) inputs and outputs. In addition, a common center-of-gravity layout and cross-coupling of the differential pair were implemented, minimizing direct-current (DC) offset and conversion between the even and odd modes without compromising RF performance. The MZM bias and driver bias currents are digitally programmed for all channels via a common control block.

Before exiting the integrated chip, the light pulses are attenuated through two VOAs: one consists of a balanced-MZI with heaters in both arms in order to tune the MZI transfer function closer to  a point of minimal transmission, while the other one is based on carrier absorption (the same as in the imb-MZI). Monitoring photodiodes have been placed at the outputs of the IM and the VOA-MZI. The total loss of the chip is around 25~dB. For testing purposes it is possible to use an alternative optical input path which is directly connected to the IM, bypassing the imb-MZI. This input has around 20~dB loss. Note that, as opposed to the receiver, loss is not an issue for the transmitter.

\section{Integrated Receiver}



On the receiver side, the integrated chip is completely passive. According to our protocol, we require its polarization independence, meaning that the visibility of the integrated receiver interferometer should be high (100$\%$ ideally) for any incoming polarization state.
We characterize the polarization independence by measuring the maximum and minimum visibilities depending on the incoming polarization state.
Additionally, the first beam splitter should also be independent of the polarization. The former requirement is difficult to achieve in PICs due to the intrinsic birefringence of the waveguides \cite{Dai:04, dai2013polarization, chang2018polarization}, which is hard to control in an imbalanced MZI. To our knowledge, only recently, a polarization independent receiver chip of a QKD system has been demonstrated \cite{Wu_recent_pol, Li:21_recent_pol}. However, the receiver in \cite{Wu_recent_pol} showed a low maximum visibility (< 98\%) and high insertion losses (excess loss up to 6~dB) and the receiver in \cite{Li:21_recent_pol} a maximum visibility of 98.7\%. In addition, a hybrid receiver based on a Michelson imbalanced interferometer and Faraday mirrors glued to the exterior of the chip has been recently validated \cite{Zhang:21}.

In the present experiment we make use of a polarization independent PIC, produced by the femtosecond laser micromachining technique \cite{Corrielli_2021}. Waveguides with low propagation losses (<~0.2~dB/cm) and low birefringence (<~$3 \cdot10^{-5}$) were inscribed in an aluminum borosilicate glass (EAGLE XG, Corning Inc.). Polarization independency of the directional couplers was achieved by exploiting the multiscan inscription technique, followed by a thermal annealing process, as described in \cite{Corrielli:18}. Furthermore, at room temperature, a careful control of the waveguides' birefringence, by fabricating compensation tracks around the waveguide of the longer arm of the imb-MZI \cite{Fernandes:12,Corrielli:18}, as well as by finely tuning the temperature of the chip, allowed for the same polarization rotation in both arms. We achieve temperature stabilization using, as on the transmitter side, a Peltier cooler/heater. At ambient temperature (around 20$^{\circ}$C) as good as perfect birefringence compensation occurs, giving rise to a minimum visibility as high as 98.9$\%$. It is important to note that this is the visibility corresponding to the case of the most unfavorable input polarization state, hence the average visibility is higher. To compare our results with the values provided above in other implementations, our maximum visibility is 99.7\%. The additional loss in the longer arm is compensated by adjusting the coupling ratio of the first coupler of the imb-MZI (about 55/45). The relationship of the visibility and QBER$_x$ is given by: $\text{QBER}_x = (1-V)/2$ and so, at the optimum temperature, its contribution to the QBER is minor. 

\begin{figure}[htbp]
\centering
\includegraphics[width=0.8\linewidth]{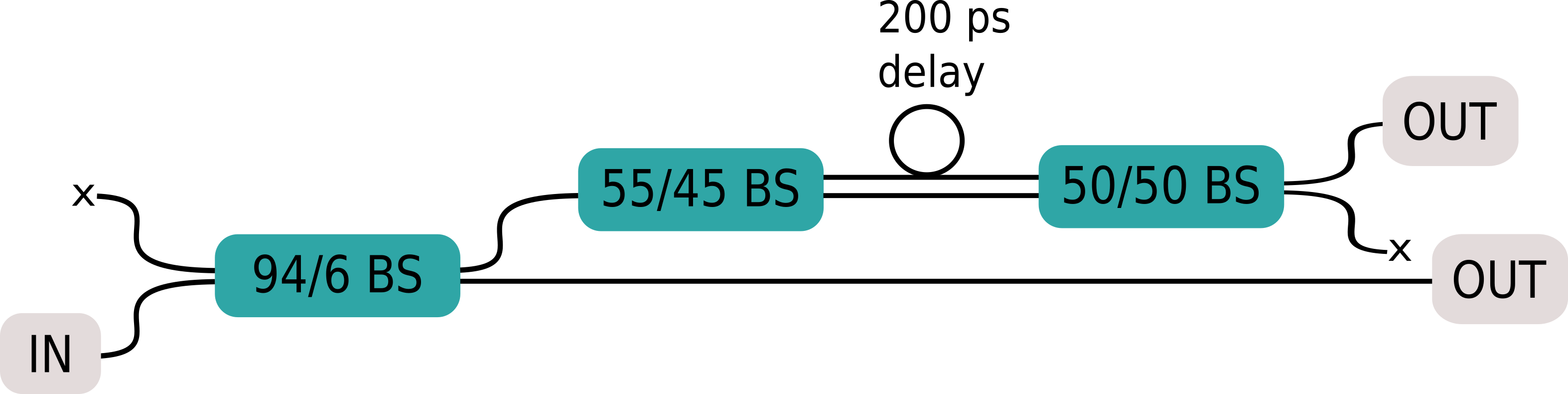}
\caption{Structure of the receiver integrated circuit.}\label{Structure_Bob}
\end{figure}

Figure \ref{Structure_Bob} shows a scheme of the receiver device. When entering the PIC, the light passes first through a 94/6 beam splitter. The majority of the light passes straight through the chip and out to a single-photon detector (SPD). The lesser amount of light goes to the imb-MZI where another SPD at one of the outputs of the interferometer detects the exiting light. The footprint of Bob is around 6~cm $\times$ 8~cm. The total loss of the chip is notably low, something that is much desired on the receiver side. In fact, we measure the excess losses for the Z and X bases, using a low-coherent light source, to be around 2.75~dB and 3.50~dB, respectively. This is excluding the splitting ratios of the first and last beam splitters, but including input/output coupling. 


\section{Results}

We performed complete secret key exchanges for different emulated distances and also employing standard SMF, using first the SNSPDs and then the InGaAs SPADs. We apply real-time error correction using a Cascade algorithm with a block size of 8192 bits \cite{cascadeJesus}. After 1000 error correction blocks, privacy amplification is executed. Thus, the total privacy amplification block size is $8.192 \cdot 10^6$ bits. In order to calculate the obtained SKR we follow the security analysis of the 1-decoy state protocol \cite{Rusca_2018} where the SKR per privacy amplification block is given by: 

$$SKR =  \frac{1}{t} [s_{0} + s_{1}(1-h(\phi_z)) - \lambda - 6 \log_2(19/\epsilon_{sec}) - \log_2(2/\epsilon_{corr})],$$

where $t$ is the block acquisition time, $s_{0}$ is the lower bound on the number of vacuum events in the Z basis and $s_{1}$ that of the single-photon events, $h(\cdot)$ is the binary entropy, $\phi_z$ is the upper bound on the phase error rate, $\lambda$ the leakage of the bits during the error correction process and $\epsilon_{sec} = 10^{-9}$ and $\epsilon_{corr} = 10^{-9}$ are the secrecy and correctness parameters, respectively.

The first set of measurements was done with the main aim to understand the maximum performance of the integrated QKD system, hence we employed the SNSPDs (see section \ref{exp_setup}). In table \ref{Table_SNSPD}, we present the results obtained using different emulated fiber distances and using a 202.0~km long single-mode fiber. The emulated fiber distances were realized using an external VOA. 

\begin{table}[H]
\begin{tabular}{ |c|c|c|c|c|c|c|c| }
\hline
 Length [km] & Attenuation [dB] & Block time [s] & RKR [kbps] & QBER$_z$ [\%] & $\phi_z$ [\%] & SKR [kbps] \\ \hline
 - & 30 & 37 & 216 & 0.9 & 1.0 & 91.0*\\ 
 
 - & 36 & 124 & 66 & 0.8 &  1.1 & 28.3\\
 
 - & 38 & 168 & 42 & 0.8 &  1.4 & 17.2\\
 
 - & 40 & 306 & 27 & 0.8 &  2.1 & 10.6\\
 \hline
 202.0 & 39.5 & 351 & 25 & 0.9 &  2.2 & 9.4\\
 \hline
 \cite{Boaron2018}: 251.7 & 42.7 & 720 & 12 & 0.5 & 2.2 & 4.9\\
 \hline
\end{tabular}
\caption{Parameters and results of secret key exchanges when using SNSPDs. * signifies estimated SKR from raw data. For comparison, the last line presents data from reference\,\cite{Boaron2018} which used a fiber-based setup with SNSPDs.
\label{Table_SNSPD}}
\end{table}


At 30~dB attenuation the number of raw detections were too high for the real-time Cascade error correction to be performed (this problem could be overcome by implementing a low-density parity check error correction on the FPGA \cite{Fadri_High_Speed}). Extremely low QBER$_z$ values for all measurements with the SNSPDs were recorded. The main contribution to the QBER$_z$ is estimated to come from the timing jitter of the SNSPD (see section \ref{exp_setup}). A small contribution to the QBER$_z$ could also come from the extinction ratio of the IM. In a static mode it is above 40~dB and it is estimated to be slightly lower in an active mode. Regarding the phase error rate, $\phi_z$, it will depend on the visibilities of the interferometers at Alice's and Bob's and the active phase stabilization between them. Thanks to  the high visibilities, $\phi_z$ is noticeably low. This is the case especially for the 30~dB attenuation due to the large number of counts, giving rise to a high raw key rate (RKR), and therefore a significant secret key rate (SKR). At higher attenuations, $\phi_z$ increases due to the lower number of counts in the X basis detector, making it harder to stabilize the phase, for further discussion see section \ref{exp_setup}. For the measurement using 202.0~km of standard SMF placed in between the transmitter and the receiver, active time-tracking was performed in order to compensate for length fluctuations in the fiber. 

It is interesting to note how the integrated version of the 3-state BB84 protocol compares with a similar fiber-based setup employing SNSPDs, described in reference \cite{Boaron2018}. Its performance with 251.7~km of ultra low-loss SMF is shown in the last line of table \ref{Table_SNSPD}. It can be concluded that with similar mean photon numbers, the same block size and around 3~dB less attenuation than the measurement performed in the fiber-based setup with 251.7~km of standard SMF, the integrated setup is practically as good as its fiber-based counterpart in terms of performance. However in terms of practicality and cost, the integrated setup is more attractive. 

In the following, we present measurements using the  practical SPADs. On one hand, these detectors are considered more qualified than the SNSPDs for industrial implementations as they are uncomplicated to cool down. On the other hand, they present higher dark count rates, after-pulsing probabilities and timing jitters, as well as lower efficiencies. The results obtained using the InGaAs SPADs are shown in table \ref{Table_InGaAs}. 

\begin{table}[H]
\begin{tabular}{ |c|c|c|c|c|c|c|c|c| }
\hline
Length & Attenuation & Dead time & Temperature & Block time & RKR & QBER$_z$ & $\phi_z$ & SKR \\ 

[km] & [dB] & [$\mu$s] & [K] & [s] & [kbps] & [\%] & [\%] & [kbps] \\ \hline

- & 30 & 20 & 188 & 453 & 18.0 & 3.6 & 2.1 & 2.9\\ 

- & 35 & 32 & 182 & 858 & 9.6 & 3.1 &  4.5 & 1.3\\

- & 40 & 20 & 188 & 1590 & 4.0 & 4.4 &  6.0 & 0.2\\
 \hline
151.5 & 29.7 & 40 & 188 & 716 & 11.0 & 3.3 &  2.7 & 1.3\\
 \hline
\cite{Boaron_2018_fast_simple}: 151.6 & 30.2 & 19 & 183 & 360 & 22.8 & 3.2 & 2.1 & 7.2 \\
\hline
\end{tabular}
\caption{Parameters and results of secret key exchanges when using InGaAs detectors. For comparison, the last line presents data of the fiber-based setup using also InGaAs detectors \cite{Boaron_2018_fast_simple}.}\label{Table_InGaAs}
\end{table}


Similar conclusions as for the results of table \ref{Table_SNSPD} can be drawn. Compared to the results with the SNSPDs, a lower RKR is observed, which is reasonable as the detector efficiency is around 20\% (a fourth of the efficiency of the SNSPDs). The increased values of QBER$_z$ are due to the higher timing jitters and afterpulsing probabilities of the InGaAs SPADs. The non-optimal 94/6 splitting ratio generates a faster saturation of detections in the Z basis, hence the consecutive increase in QBER$_z$ as well as non-negligible dark count rates for higher attenuations in the X basis. 151.5~km standard SMF was also placed in between the transmitter and the receiver. Due to the lower number of counts and therefore increased difficulty to perform perfect time-tracking and active phase tracking (see section \ref{exp_setup}), $\phi_z$ is slightly higher than its attenuated analogue.

Again we compare these results with those obtained using the same detectors and protocol in a fiber-based setup, more precisely, the one in reference \cite{Boaron_2018_fast_simple}. At a distance of 151.6~km, with half of the mean photon numbers and the same block size, the fiber-based setup seems to perform better in terms of RKR and SKR than the integrated one with these detectors; however, this difference can be attributed mainly to the fact that the detectors were operated with different parameters. In fact, the fixed, yet non-optimal, splitting ratio at the receiver side of the integrated QKD setup forced a lower bias voltage and higher dead time in the X basis to minimize the dark counts (while lowering the detector efficiency) and maximise the number of counts, respectively. However, the comparable values on QBER$_z$ and $\phi_z$ makes the employment of the integrated devices still attractive. In particular, the replacement of the first beam splitter with a tunable MZI, a device already well optimized on the same platform \cite{Albiero2022}, will allow for an optimal splitting ratio at the receiver side with a negligible cost in terms of losses and device complexity.

\begin{figure}[H]
\centering
\includegraphics[width=0.8\linewidth]{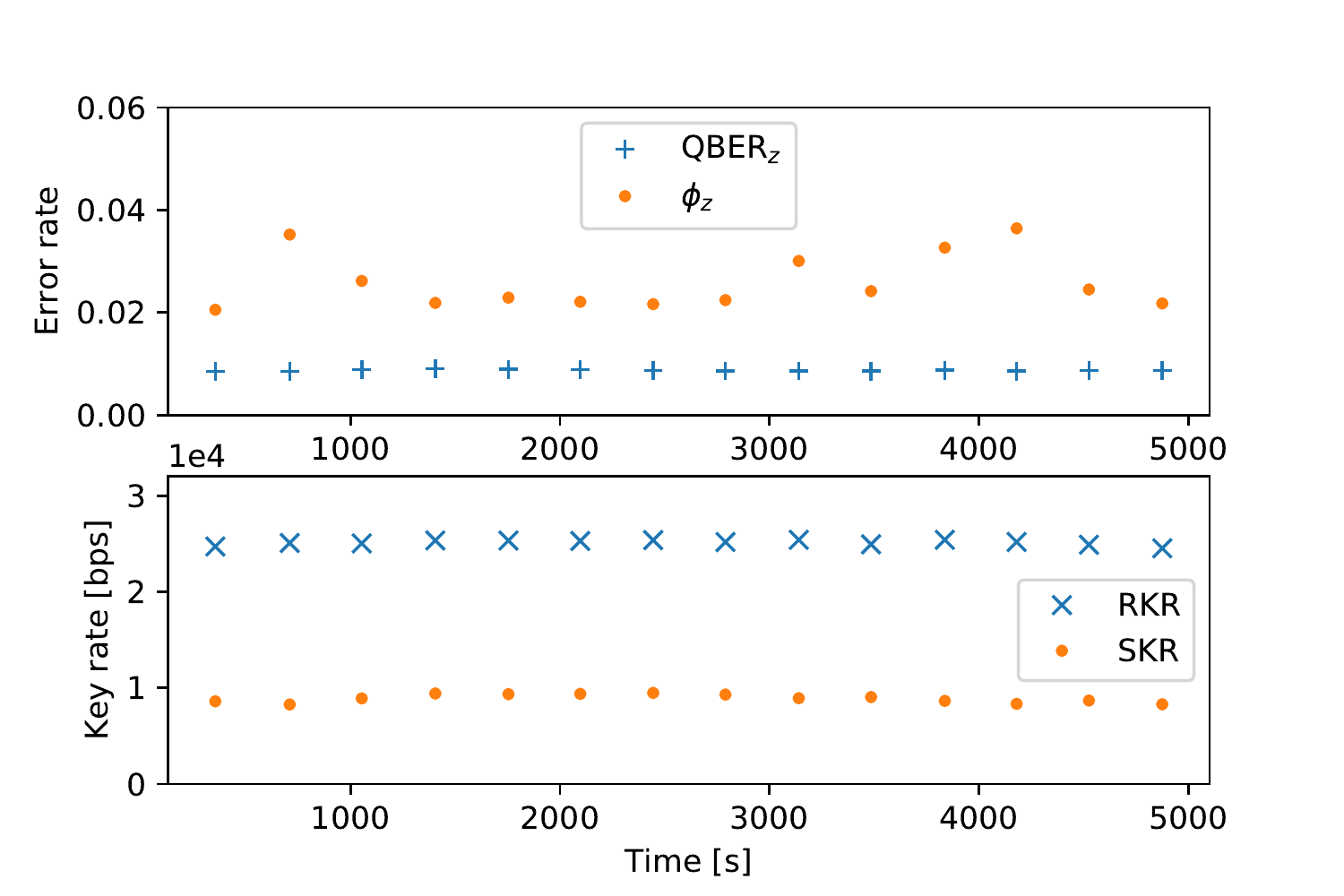}
\caption{QBER$_z$, $\phi_z$, RKR and SKR during several secret key exchanges over 80~min.}\label{total_results_snspd_200km}
\end{figure}

Lastly, we present the complete results of the integrated QKD setup with 202.0~km of standard SMF and SNSPDs as detectors with a secret key exchange let run for around 80~min. In figure~\ref{total_results_snspd_200km} the RKR, SKR, QBER$_z$ and $\phi_z$ are shown as a function of time.
We observe a stable RKR and SKR, around 25~kbps and 9~kbps, respectively. The same goes for the QBER$_z$, around 0.9~\%, thanks to the high number of detections in the Z basis and so an excellent time-tracking. Concerning $\phi_z$, as previously mentioned, more fluctuations are observed due to a lower detection rate in the X basis and so a more complicated time-tracking and active phase adjustment (refer to section \ref{exp_setup}). 

\section{Conclusion}
An integrated QKD system has been presented and shown to perform as good as its fiber-based analogue, and most importantly as the state-of-the-art of integrated QKD systems \cite{Liu_review}. Its transmitter is practical and low-cost thanks to the integration of the imb-MZI and, especially, the IM and corresponding electronics. Additionally, its receiver features low loss and is polarization independent, which is typically complicated to achieve in integrated platforms.


Even though polarization fluctuations of QKD systems are nowadays very well controlled and compensated in laboratory conditions \cite{Peng, Xavier:08}, it might still be demanding to compensate for particularly rapid fluctuations in polarization that could occur in real-world fiber-optic lines, e.g. because of trains passing or lightning strikes \cite{Exfo}. Thus, the integrated QKD system here suggested, based on time-bin encoding and polarization insensitivity, testifies for effortless integration in present-day fiber-optic networks. 



We believe that the integrated high-speed QKD system gives an important contribution to the advancement of integrated quantum technologies and simultaneously reflects their maturity. Future investigations could cover how to integrate all components on-chip (meaning the laser on the transmitter side and the SPDs on the receiver side), which has the risk of being costly due to the active materials required, such as InP, and further complicated due to the need of interfacing different active and non-active materials via gluing or bonding. Several works have already examined the merge of InP platforms with silicon platforms \cite{Roelkens, Bowers_liang}. On the transmitter side of the present integrated platform, the PIC, the driver EIC and all DC control loops could be monolithically integrated in a single electronic and photonic IC (EPIC) chip. The EPIC technology \cite{EPIC_Hanjo} for this approach is mature and already in use for data center applications. An adaptation to QKD applications is only a matter of chip design rather than process development. Furthermore, EPIC and even PIC/EIC solutions can be scaled to significantly higher modulation rates, however limited by the achievable extinction ratio. Thanks to the small dimensions of the introduced integrated platforms, it is rather straight-forward to integrate the current QKD system in two 19-inch racks, ready for usage in a real-work network. 

\begin{backmatter}
\bmsection{Funding}
We thank the Eurostars project E!11493 QuPIC for financial support. SA and AC acknowledge funding by PRIN2017 programme of the Italian Ministry for University and Research,  QUSHIP project (Id. 2017SRNBRK). RO acknowledges funding by the European Union through the ERC Advanced Grant CAPABLE (742745).

\bmsection{Acknowledgments}
We thank Claudio Barreiro for providing the electronic cards and Federico Bassi for his preliminary work on the fabrication and characterization of the receiver.

\bmsection{Disclosures}
The authors declare no conflicts of interest.

\bmsection{Data Availability}
Data underlying the results presented in this paper are not publicly available at this time but may be obtained from the authors upon reasonable request.
\end{backmatter}


\bibliography{bibliography}

\end{document}